\documentclass{article}
\usepackage{ijcai22}

\usepackage{times}
\usepackage{soul}
\usepackage{url}
\usepackage[hidelinks]{hyperref}
\usepackage[utf8]{inputenc}
\usepackage[small]{caption}
\usepackage{graphicx}
\usepackage{amsmath}
\usepackage{amsthm}
\usepackage{booktabs}
\usepackage{algorithm}
\usepackage{algorithmic}

\usepackage[table,xcdraw]{xcolor}
\usepackage{diagbox}
\usepackage{amssymb} 
\usepackage{multirow}
\usepackage{xcolor}
\usepackage{xspace} 
\usepackage{color} 
\usepackage{arydshln} 
\usepackage{bm}
\newcommand{\cert}{\textsc{CERT}\xspace}
\newcommand{\certg}{\textsc{CERT}g\xspace}
\newcommand{\peval}{PandasEval\xspace}
\newcommand{\neval}{NumpyEval\xspace}
\newcommand{\codepy}{\textsc{PyCodeGPT}\xspace}
\newcommand{\codepyxl}{\textsc{PyCodeGPT}-XL\xspace}
\newcommand{\codegen}{\textsc{CodeGen}\xspace}
\newcommand{\codegenxl}{\textsc{CodeGen}-XL\xspace}
\newcommand{\pandas}{Pandas\xspace}
\newcommand{\numpy}{NumPy\xspace}
\newcommand{\pytorch}{PyTorch\xspace}
\newcommand{\citet}[1] {\citeauthor{#1}~\shortcite{#1}} 

\title{\cert: Continual Pre-Training on Sketches for Library-Oriented Code Generation}

\author{
Daoguang Zan$^{1,2*}$\and
Bei Chen$^3$\and
Dejian Yang$^3$\and
Zeqi Lin$^3$\and
Minsu Kim$^{4}$\thanks{Work done during the internship at Microsoft Research Asia.}\and\\
Bei Guan$^{2,5}$\and
Yongji Wang$^{2,5,6}$\and
Weizhu Chen$^7$\and
Jian-Guang Lou$^{3}$
\affiliations
$^1$Cooperative Innovation Center, Institute of Software, Chinese Academy of Sciences\\
$^2$University of Chinese Academy of Sciences\\ 
$^3$Microsoft Research Asia\\ 
$^4$Korea University\\
$^5$Integrative Innovation Center, Institute of Software, Chinese Academy of Sciences\\
$^6$State Key Laboratory of Computer Science, Institute of Software, Chinese Academy of Sciences\\
$^7$Microsoft Azure AI
\emails
\{daoguang@, guanbei@, ywang@itechs.\}iscas.ac.cn; minsu@korea.ac.kr\\
\{beichen, deyang, zeqi.lin, wzchen, jlou\}@microsoft.com
}

\begin{document}

\maketitle

\begin{abstract}

Code generation is a longstanding challenge, aiming to generate a code snippet based on a natural language description. Usually, expensive text-code paired data is essential for training a code generation model. Recently, thanks to the success of pre-training techniques, large language models are trained on large-scale unlabelled code corpora and perform well in code generation. In this paper, we investigate how to leverage an unlabelled code corpus to train a model for library-oriented code generation. Since it is a common practice for programmers to reuse third-party libraries, in which case the text-code paired data are harder to obtain due to the huge number of libraries. We observe that library-oriented code snippets are more likely to share similar code sketches. Hence, we present \cert with two steps: a sketcher generates the sketch, then a generator fills the details in the sketch. Both the sketcher and the generator are continually pre-trained upon a base model using unlabelled data. Furthermore, we craft two benchmarks named \peval and \neval to evaluate library-oriented code generation. Experimental results demonstrate the impressive performance of \cert. For example, it surpasses the base model by an absolute $15.67\%$ improvement in terms of pass$@1$ on \peval. Our work is available at \url{https://github.com/microsoft/PyCodeGPT}.

\end{abstract}

\begin{figure}[t]
    \small
    \centering
    \includegraphics[width=0.80\linewidth]{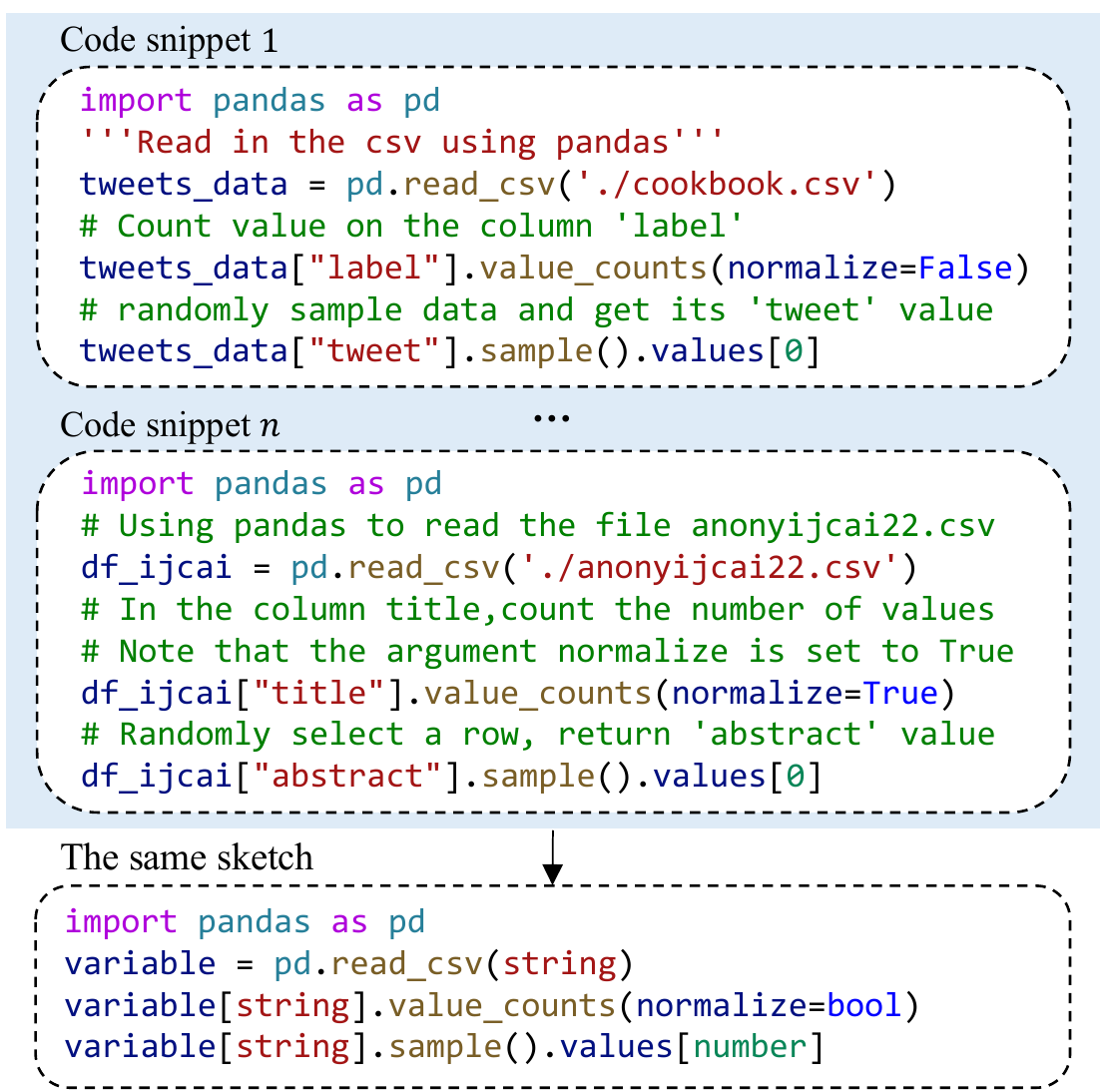}
    \caption{ An example in Python: multiple code snippets using \pandas may have the same sketch after anonymizing user-defined terms.
    }
    \label{fig:figure1}
\end{figure}

\section{Introduction} \label{introduction}

Code generation, aiming to generate a code snippet for a given natural language description, is a longstanding challenge in the artificial intelligence community. Usually, to train a code generation model with good performance, the massive amount of code snippets paired with natural language descriptions are indispensable~\cite{sun2019grammar,lu2021codexglue}. However, it is costly and time-consuming to annotate such a dataset. To alleviate this problem, inspired by GPT-3's powerful zero-shot natural language generation ability~\cite{brown2020language}, recent years have witnessed a trend to train large language models using large-scale code corpora (e.g., GitHub), and expect these models to work well directly on code generation tasks, without fine-tuning on expensive text-code pairs.
For example, Codex shows that a $12$B parameters language model can solve $28.8\%$ of standalone Python programming problems\footnote{It is measured on HumanEval~\cite{chen2021evaluating} with pass$@1$.}.

In this paper, we focus on investigating whether and how language models pre-trained on code corpora (without fine-tuned on pairwise labelled data) can generate library-oriented code snippets rather than standalone ones.
During software development, it is a common practice for programmers to reuse third-party libraries (e.g., \pandas and \numpy) to implement needed functionalities.
It is not easy for programmers to learn how to use these libraries properly. For example, according to our statistics, more than $40\%$ of StackOverflow questions with ``Python'' tag also have at least one library tag.
Moreover, for library-oriented code generation, the necessity of training the model without pairwise labelled data is raised, as programmers usually need to reuse different libraries in different scenarios, and it is extremely costly to label sufficient text-code pairs that cover most of these libraries.

Compared to standalone code snippets, library-oriented code snippets are more likely to share similar sketches.
\emph{Sketch} is the code structure after anonymizing the user-defined terms in the code, such as variable names, method names, constants, etc., which has also been identified as an API usage pattern in previous research litterateurs on software data mining~\cite{ZhongECOOP2009,Wang2013-MSR,NIU2017127}.
An example is shown in Figure~\ref{fig:figure1}. After anonymizing variables and constants, multiple code snippets using the \pandas APIs may have the same (or similar) sketch.
Based on this observation, a natural idea to improve library-oriented code generation is to decompose this task into two subtasks: generating the sketch and then filling in the details.
Many methods based on this idea have been proposed in different code generation tasks (e.g., Coarse-to-Fine~\cite{dong2018coarse} and \textsc{PlotCoder}~\cite{chen2021plotcoder}) and have shown that this idea can effectively improve the quality of generated code snippets.
However, these methods are proposed for the fine-tuning process, in which high-quality text-code pairs are required to derive supervision signals for the two-step generation.
Therefore, in our scenario that no pairwise labelled data is provided, a research question arises: how to leverage the insight of sketching to enhance the language model pre-training on unlabelled code corpora, thus improving the quality of generated library-oriented code snippets?

To meet the challenge, we propose \cert (for sket\textbf{C}her and  g\textbf{E}ne\textbf{R}a\textbf{T}or), a continual pre-training approach on sketches for library-oriented code generation. In \cert, a sketcher firstly focuses on predicting a sketch, which omits user-defined details; then, a generator uses the sketch as a prompt to generate the complete code. Both the sketcher and the generator are continually pre-trained based on a base language model for code, using unlabelled code corpora rather than pairwise labelled data. In addition, we craft two evaluation benchmarks for Python libraries, called \peval and \neval,  each including $101$ programming problems using \pandas and \numpy, respectively. 
We perform extensive experiments on \cert. 
Results indicate that \cert has superior performance on library-oriented code generation.
We further draw several insights via thorough analysis. 

\begin{figure}[t] 
    \small
    \centering
    \includegraphics[width=0.85\linewidth]{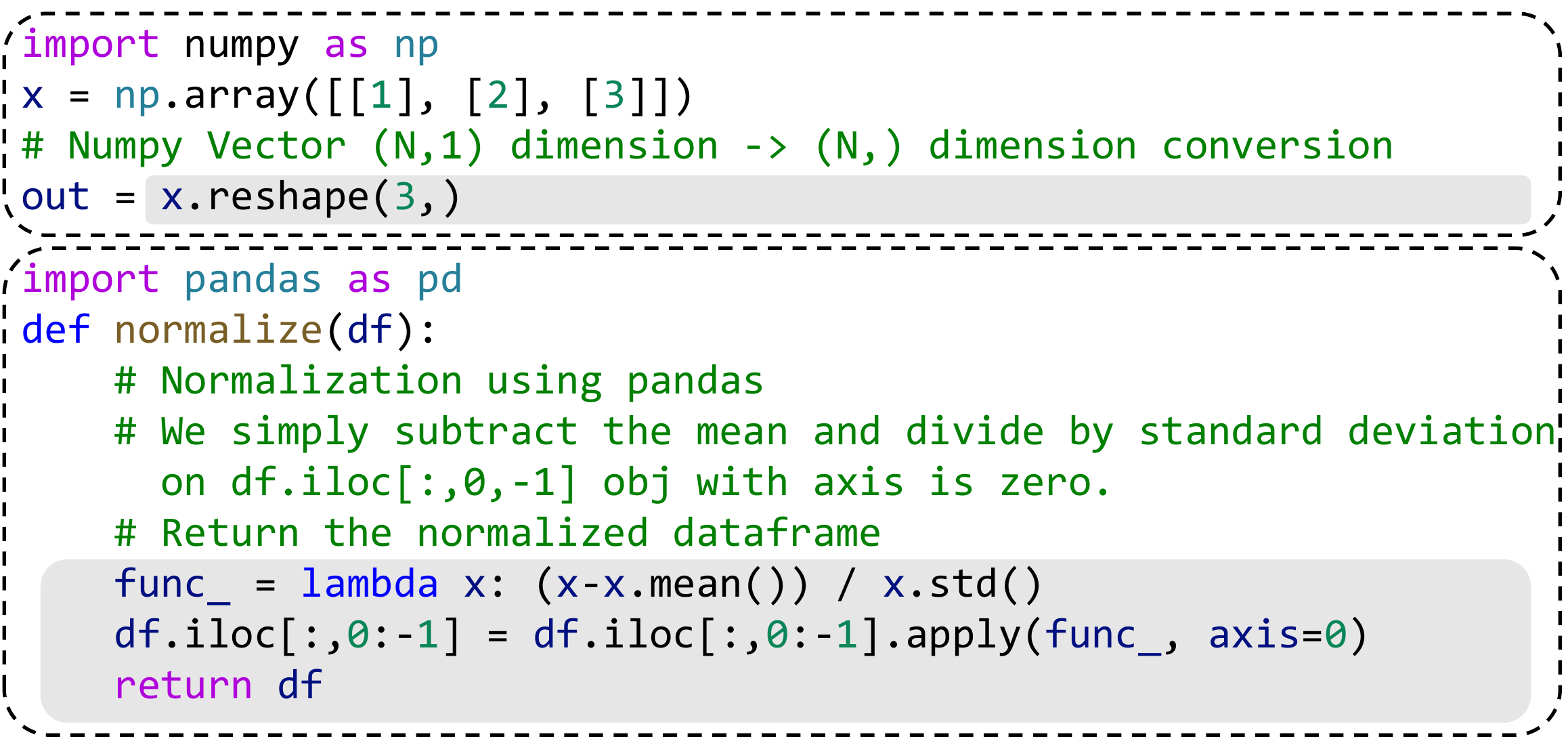}
    \caption{Two examples of programming problems from the \peval and \neval benchmarks. Context is shown with a white background and the target code with a gray background.}
    \label{fig:examples}
\end{figure}

\section{Task Formulation}\label{sec:formulation}
Before diving into the details of our proposed approach, we start with a formal description of the task.
Code generation is to solve a programming problem: generate \emph{target code} based on \emph{context}. Context contains natural language problem description in the form of code comments, and a code snippet that includes statements such as import, function header and variable definition; target code is a code snippet that solves the programming problem described in the context. Formally, let $\mathbf{x}=(x_1,x_2,{\cdots},x_N)$ denote the context, where each $x_n$ can be either a code token or a natural language token. Given $\mathbf{x}$, the code generation model can be formulated as $\mathbf{y}=\mathcal{M}(\mathbf{x})$, where $\mathbf{y}=(y_1,y_2,{\cdots},y_M)$ denotes the target code and each $y_m$ is a code token.

For standalone code generation, the programming problem is expected to be solved by a code snippet without using third-party libraries; conversely, for library-oriented code generation, the target code $\mathbf{y}$ contains library API calls. Two examples of library-oriented programming problems can be found in Figure~\ref{fig:examples}. Note that carefully labelled context and target code pairs are indispensable for model fine-tuning, while our proposed approach only requires continual pre-training on unlabelled code corpora.

\section{Methodology} \label{methodology}
In this section, we introduce our base models, followed by the details of our proposed approach \cert.

\subsection{Base Models} \label{basemodel}
Codex~\cite{chen2021evaluating} is a milestone pre-trained model that can generate decent code, but it is not publicly available. Several attempts have been made to reproduce Codex's powerful code generation capability, e.g.,  CodeClippy\footnote{\url{https://github.com/CodedotAl/gpt-code-clippy}} and CodeParrot\footnote{\url{https://huggingface.co/transformersbook/codeparrot}}, but their performance in Python are not satisfactory. To this end, we present \codepy~\footnote{More details about \codepy are in Appendix.}, a pre-trained language model, which has the ability to generate pretty good standalone Python code, for example, achieving $8.33\%$ pass$@1$ on HumanEval~\cite{chen2021evaluating}. 
Specially, \codepy is a $110$M parameters model based on GPT-Neo~\cite{gpt-neo}. We collected $60.6$M raw python files with a total size of $330$GB. After a series of data pre-processing strategies, such as de-duplicating python files, cleaning and formatting the contents, etc., the final pre-training corpus contains about $13.0$M high-quality python files with the size of $96$GB. 
\codepy is pre-trained for $200$K steps and $100$B tokens on a cluster of $16$ NVIDIA V$100$ GPUs with $32$GB memory. The pre-training time is about $2$ days.
We summarize the three key points that make \codepy powerful: 1) a large amount of carefully cleaned data for pre-training; 2) a newly trained tokenizer, which is specialized in python; and 3) a resampling strategy that prioritizes high-quality data.
Besides \codepy, we also regard \codegen (\textsc{Mono} $350$M)~\cite{nijkamp2022conversational} as one of our base models, which is by far the best performing publicly available model on HumanEval\footnote{\codegen was released during the review period of this paper.}.

\begin{figure}[t]
    \small
    \centering
    \includegraphics[width=0.80\linewidth]{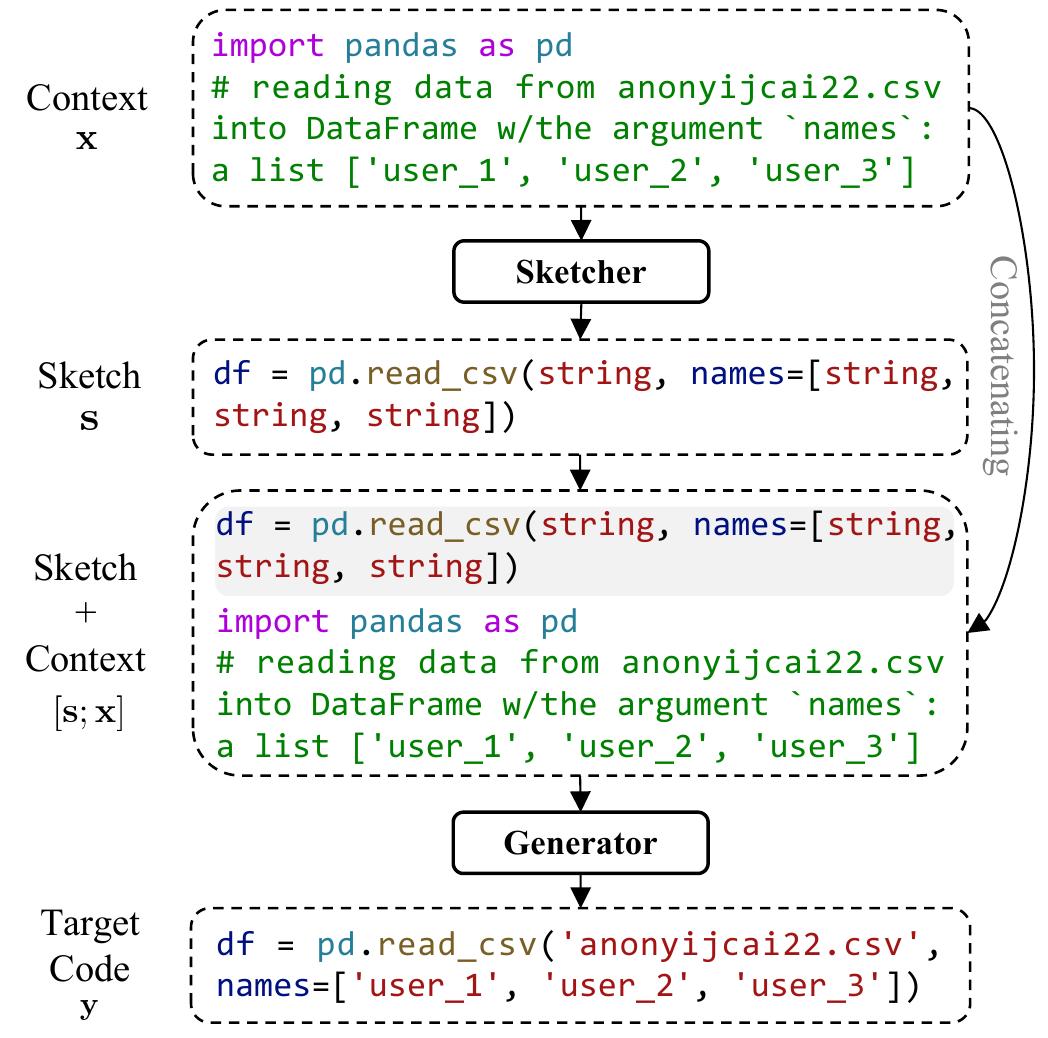}
    \caption{Overview of \cert: a sketcher and a generator.}
    \label{fig:figure2}
\end{figure}

\subsection{\cert}\label{cert}
As mentioned in Section~\ref{sec:formulation}, code generation is to generate target code $\mathbf{y}$ based on context $\mathbf{x}$. 
Since we observe that library-oriented code snippets are more likely to share similar sketches, we present a novel approach \cert and decompose the code generation model $\mathcal{M}$ into two modules: a sketcher $\mathcal{M}_{\rm{S}}$ and a generator $\mathcal{M}_{\rm{G}}$.
Figure~\ref{fig:figure2} shows the overview of \cert with a concrete example in \pandas. Given $\mathbf{x}$ as the input, the sketcher predicts $\mathbf{s}$, which is the sketch of the target code $\mathbf{y}$. The sketcher generates multiple candidate sketches ($200$ in our experiments) and we choose the one that appears the most. Then, the input of the generator is the concatenation of $\mathbf{s}$ and $\mathbf{x}$. Formally, the process of \cert can be written as $\mathbf{s} =\mathcal{M}_{\rm{S}}(\mathbf{x})$ and $\mathbf{y} = \mathcal{M}_{\rm{G}}([\mathbf{s}; \mathbf{x}])$.
Note that if the sketch $\mathbf{s}$ is already a complete code snippet without anonymous symbols, we directly take it as the final prediction instead of using the generator; and if the sketch $\mathbf{s}$ is an empty sequence, we directly feed $\mathbf{x}$ into the generator. 

We build the sketcher and generator on the top of the base model (\codepy or \codegen) by continual pre-training. At first, we extract the python files that use a specific library (e.g., \pandas) from the whole pre-training corpus ($13.0$M files mentioned in Section~\ref{basemodel}), and obtain the sub-corpus denoted by $\mathcal{D}$. 
Then, we will detail the continual pre-training process of the sketcher and generator for this library.

\begin{figure}[t]
    \small
    \centering
    \includegraphics[width=0.92\linewidth]{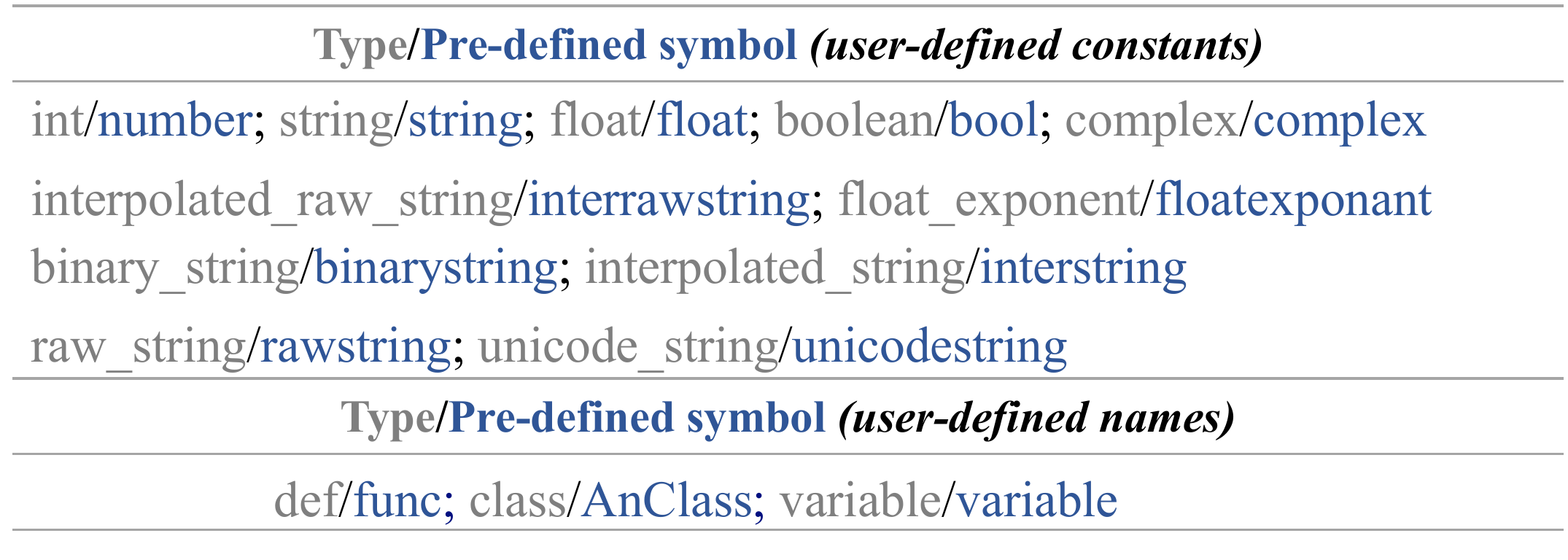}
    \caption{The pre-defined symbols in sketcher.}
    \label{fig:figure8}
\end{figure}

\begin{figure*}[t]
    \small
    \centering
    \includegraphics[width=0.78\linewidth]{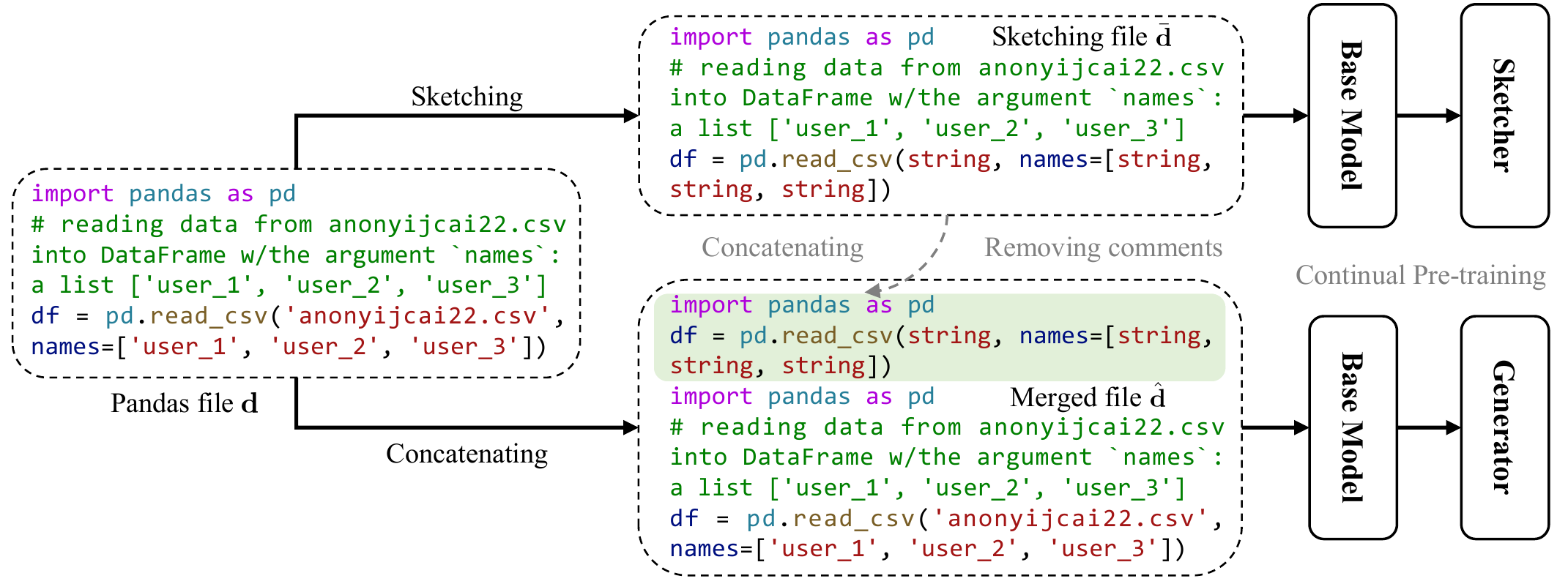}
    \caption{Training data preparation for sketcher and generator with an example in \pandas.}
    \label{fig:figure3}
\end{figure*}

\paragraph{Sketcher.}
Given the library-oriented sub-corpus $\mathcal{D}$, we perform the sketching operation on each file $\mathbf{d}\in\mathcal{D}$. An example is shown in the upper part of Figure~\ref{fig:figure3}. The sketching operation is used to anonymize the user-defined terms in the code file with our pre-defined symbols. The file after sketching is denoted as $\bar{\mathbf{d}}$. We design three different types of sketching operations: 1) only anonymizing the user-defined constants (Default \cert); 2) only anonymizing the user-defined names, including function names, class names, and variable names (\cert-N); and 3) anonymizing both the user-defined constants and names (\cert-NC). For example, in Figure~\ref{fig:figure3}, the constant `\texttt{user\_1}' is anonymized with the pre-defined symbol `\texttt{string}'. The details of pre-defined symbols are shown in Figure~\ref{fig:figure8}. Then, we continually pre-train the base model on the library-oriented corpus after sketching, and we obtain the sketcher model. The pre-training objective is the same as that of the base model. We pre-train the model for $100$K steps on a cluster of $8$ NVIDIA V$100$ GPUs with $32$GB memory. 

\paragraph{Generator.}
In order to prepare the pre-trained corpus for the generator, we firstly split the original file $\mathbf{d}$ and the sketching file $\bar{\mathbf{d}}$ into $K$ blocks\footnote{We use pip-tools: autopep8, docformatter and redbaron.}, and obtain $\mathbf{d}=(d_1,d_2,{\cdots},d_K)$ and $\bar{\mathbf{d}}=(\bar{d}_1,\bar{d}_2,{\cdots},\bar{d}_K)$. Each block is a relatively complete code snippet, such as a function or a class. Note that before splitting, we remove the natural language code comments from the sketching file $\bar{\mathbf{d}}$. Then, the two files are cross-merged to give a merged file $\hat{\mathbf{d}}=(\bar{d}_1, d_1,\bar{d}_2,d_2,{\cdots},\bar{d}_K,d_K)$. This is to mimic the process of having a sketch as a prompt for each block. An example is shown in the lower part of Figure~\ref{fig:figure3}. Then, the base model is continually pre-trained on all the merged files $\hat{\mathbf{d}}$ and we obtain the generator model. As with the sketcher model, we continually pre-train for $100$K steps.

\section{Benchmark Construction} \label{pandasevalandnumpyevaldatasets}

Third-party libraries are widely used in reality, while little work has been done to evaluate library-oriented code generation. To meet this challenge, we craft \peval and \neval, two benchmarks for library-oriented code generation in Python. Each sample in the benchmarks is a programming problem consisting of context and target code. The programming problems are solved using libraries, where \pandas is for \peval, and \numpy is for \neval. The benchmarks are expected to be diverse, authentic, high quality, moderately difficult, and unseen during pre-training.

In order to craft programming problems using libraries, we refer to StackOverflow\footnote{\url{https://stackoverflow.com}}, a Q\&A website for programmers.
There are plenty of real-world programming problems posted by real users, which helps us to improve the authenticity of our data. 
Specifically, we search for posts using the library tag on StackOverflow, and select those with high votes. To ensure quality, we only refer to posts with accepted answers. We go through a post's question and its accepted answer, then manually organize them into the form needed for our benchmarks, containing both context and target code. We also polish all programming problems so that the problem descriptions are clear and the codes are correct. Note that we keep the intentions and the descriptions of the programming problems consistent with the posts to the maximum extent.
Finally, two programmers with more than three years of coding experience in the library are invited to act as code generation models and check the quality of the data.

As a result, we craft $101$ programming problems for \peval and \neval, respectively. 
Each programming problem is equipped with test cases for evaluation. 
For the programming problems in the form of a function, such as the bottom one in Figure~\ref{fig:examples}, we create $20$ test cases for each of them. For the others that contain no functions, such as the top one in the Figure~\ref{fig:examples}, we provide $1$ test case to check the correctness of predicted variable (e.g., \texttt{out} in Figure~\ref{fig:examples}). 
In total, $64\%$ programming problems in \peval and $30\%$ in \neval are equipped with $20$ test cases.
In addition, we craft programming problems that refer to StackOverflow rather than GitHub, and also carefully organize and polish the problems, so that we can ensure they are unseen by the pre-trained models.

\section{Experiments} \label{experiments}
In this section, we evaluate \cert on \peval and \neval to verify its effectiveness.

\paragraph{Evaluation Metrics.}
We use pass$@k$ as the metrics. When $k$ code samples are generated per problem, pass$@k$ indicates the fraction of correct ones. But computing pass$@k$ in this way may have high variance. Hence, we follow \citet{chen2021evaluating} to generate $n\ge k$ code samples per problem ($n=200$ in our experiments) and count the number of correct samples $c$. If $n-c<k$, then pass$@k=1$; otherwise, ${\rm{pass}}@k=1-\prod\nolimits_{i=n-c+1}^{n} (1-k/i)$. 
Note that a predicted code is correct if it can pass all the test cases.

\paragraph{Implementation Details.}
We implement our approach using \pytorch~\cite{paszke2017automatic},  HuggingFace's transformers library~\cite{Wolf2019HuggingFacesTS}, and DeepSpeed\footnote{\url{https://github.com/microsoft/DeepSpeed}}. 
In the training phase of \codepy, we set the batch size to $10$, the window size to $1024$, the learning rate to $5$e-$4$, the gradient accumulation steps to $4$ and the weight decay to $0.1$. 
The settings of sketcher and generator are the same as \codepy. 
We use the mixed-precision of FP$16$ to accelerate the pre-training.
In inference phase, we set the temperature to one of $[0.1, 0.2, 0.3, 0.4, 0.5, 0.6, 0.7, 0.8, 0.9, 1.0]$. The best performance is reported across the above hyper-parameters. 

\begin{table}[t]
\centering
\resizebox{75mm}{!}{
\begin{tabular}{lc|lc}
\hline
\textbf{Model}  & \textbf{pass$@1$} & \textbf{Model} & \textbf{pass$@1$} \\ \hline
GPT-Neo $125$M    & $0.75$            & GPT-Neo $1.3$B     & $4.79$            \\
AlphaCode $89$M & $4.30$            & CodeParrot $110$M    & $3.80$            \\ \hline
Codex $42$M       & $5.06$            & Codex $85$M      & $8.22$            \\
Codex $2.5$B       & $21.36$            & Codex $12$B      & $28.81$            \\
\hline
\codepy $110$M     & $8.33$            & \textsc{Code}\textsc{Gen}-\textsc{Mono} 350M               & $12.76$                \\ \hline
\end{tabular}
}
\caption{The pass$@1$ ($\%$) results on HumanEval benchmark. We omit CodeT5 ($220$M), CodeGPT-Adapted ($124$M), and CodeClippy ($125$M) as their pass$@1=0$.}
\label{tab:table2}
\end{table}
\subsection{Main Results}\label{main_results}
Before evaluating \cert, we would like to evaluate our base model \codepy on HumanEval~\cite{chen2021evaluating} compared to several advanced pre-trained models. As shown in Table~\ref{tab:table2}, \codepy ($110$M) achieves competitive $8.33\%$ pass$@1$. It largely exceeds other models with comparable parameters, e.g., AlphaCode ($89$M)~\cite{li2022competition}, CodeClippy ($125$M), and CodeParrot ($110$M), and also is better than the larger model GPT-Neo ($1.3$B).

Then, our proposed \cert is evaluated on \peval and \neval. 
We train CERT on two base models, including \codepy and \codegen, named \codepy-\cert and \codegen-\cert, respectively.
For each benchmark, we extract corresponding library-oriented files to train \cert. The file numbers are about $0.61$M for \pandas and $2.62$M for \numpy.
Baselines include our base models \codepy and \codegen; \codepyxl and \codegenxl, which are continual pre-trained \codepy and \codegen on the extracted library-oriented files; and advanced pre-trained models for code, like  CodeT5~\cite{wang2021codet5}, CodeGPT~\cite{lu2021codexglue}, CodeClippy and CodeParrot. Table~\ref{tab:table3} summarizes the performance. \cert consistently outperforms all the baselines by a large margin. The absolute improvements over \codepy and \codegen are shown in red, which are significant, for example, $12.69\%$ pass$@1$ for \codegen-\cert and $13.43\%$ pass$@1$ for \codepy-\cert on \neval. 
The results demonstrate the effectiveness of \cert with the idea of leveraging sketches for library-oriented code generation.

\begin{figure}[t]
    \small
    \centering
    \includegraphics[width=0.80\linewidth]{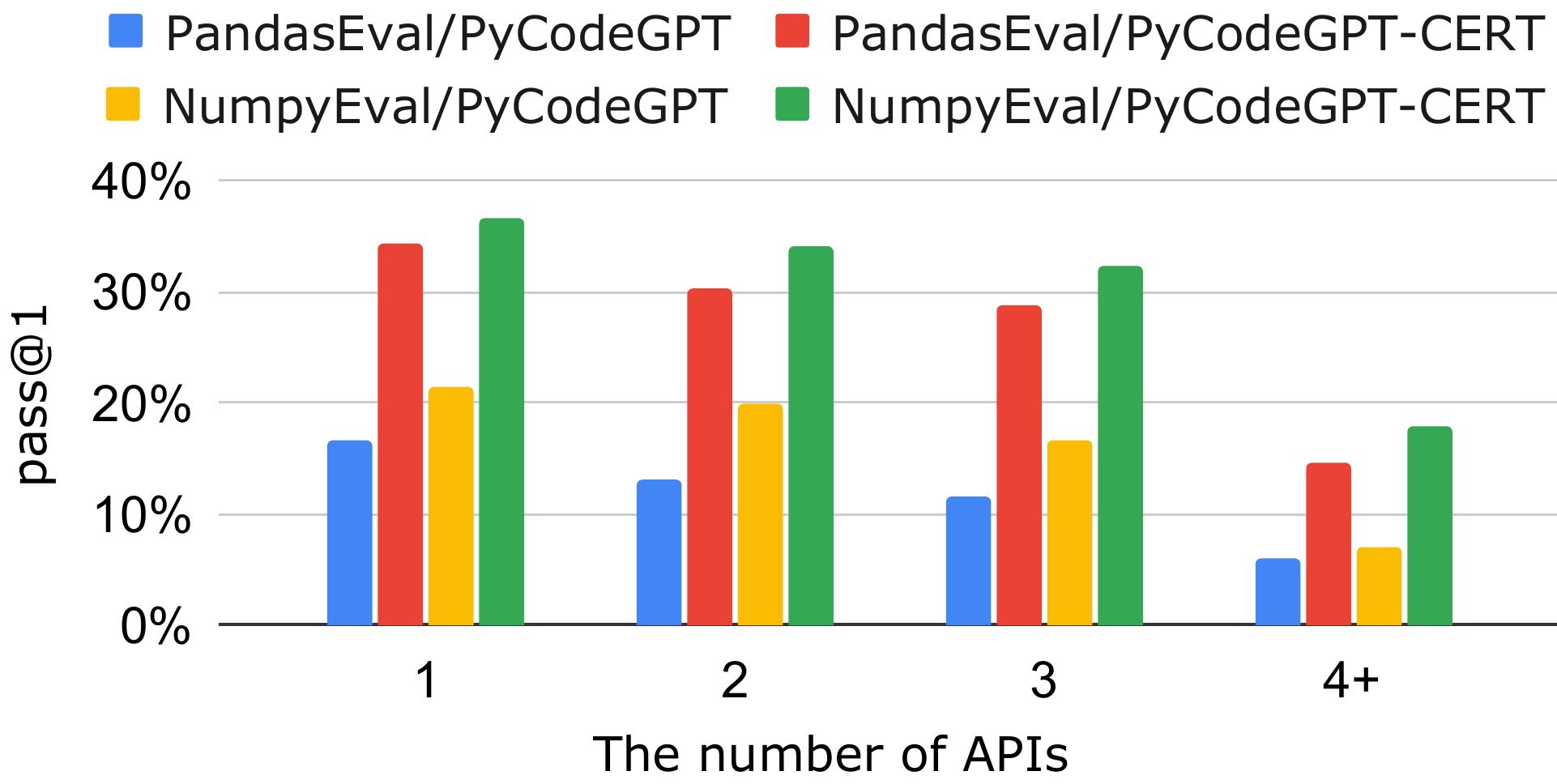}
    \caption{The pass$@1$ result with respect to the number of APIs.
    }
    \label{fig:figure10}
\end{figure}

Additionally, we would like to investigate the performance of \cert with respect to the number of API calls involved in the target code. We divided the programming problems in each benchmark into four parts based on the number of APIs. As shown in Figure~\ref{fig:figure10}, compared to \codepy, \codepy-\cert has a steady improvement on each part. It indicates that \cert can improve the performance of library-oriented code generation of varying difficulties.

\begin{table}[t]
\resizebox{86mm}{!}{
\centering
\begin{tabular}{c|l|ccc}
\hline
\multicolumn{1}{l|}{\textbf{Benchmark}} & \textbf{Model} & \textbf{pass$@1$} & \textbf{pass$@10$} & \textbf{pass$@100$} \\ \hline
\multirow{13}{*}{\begin{tabular}[c]{@{}c@{}}Pandas\\ Eval\end{tabular}} & CodeT5 ($220$M) & $0.00$ & $0.00$ & $0.00$ \\
 & CodeGPT-Adapted ($124$M) & $0.62$ & $2.65$ & $4.95$ \\
 & CodeClippy ($125$M) & $0.14$ & $0.92$ & $1.92$ \\
 & CodeParrot ($110$M) & $3.21$ & $13.62$ & $33.27$ \\
 & \codegen ($350$M) & $14.24$ & $30.71$ & $46.04$ \\
 & \codegenxl & $21.07$ & $37.67$ & $49.07$ \\
 & \codegen-\cert & \bm{$26.40$}\color{red}{$\blacktriangle12.16$} & \bm{$46.49$}\color{red}{$\blacktriangle15.78$} & \bm{$58.16$}\color{red}{$\blacktriangle12.12$} \\
 & \codepy ($110$M) & $12.75$ & $37.80$ & $59.65$ \\
 & \codepyxl & $19.80$ & $46.80$ & $60.04$ \\
 & \codepy-\cert & \bm{$28.42$}\color{red}{$\blacktriangle15.67$} & \bm{$48.04$}\color{red}{$\blacktriangle10.24$} & \bm{$60.96$}\color{red}{$\blacktriangle1.31$} \\
 & \ - \codepy-\cert-N & $23.66$ & $41.73$ & $55.08$ \\ 
 & \ - \codepy-\cert-NC & $19.07$ & $41.50$ & $54.82$ \\ 
 & \ - \codepy-\certg & $20.58$ & $42.61$ & $56.00$ \\ 
 \hline
 
\multirow{13}{*}{\begin{tabular}[c]{@{}c@{}}Numpy\\ Eval\end{tabular}} & CodeT5 ($220$M) & $0.00$ & $0.10$ & $0.74$ \\
 & CodeGPT-Adapted ($124$M) & $1.59$ & $4.17$ & $8.54$ \\
 & CodeClippy ($125$M) & $0.08$ & $0.59$ & $1.24$ \\
 & CodeParrot ($110$M) & $8.42$ & $21.46$ & $45.94$ \\
 & \codegen ($350$M) & $19.31$ & $40.89$ & $60.58$ \\
 & \codegenxl & $27.33$ & $44.75$ & $63.39$ \\
 & \codegen-\cert & \bm{$32.00$}\color{red}{$\blacktriangle12.69$} & \bm{$49.45$}\color{red}{$\blacktriangle8.56$} & \bm{$67.82$}\color{red}{$\blacktriangle7.24$} \\
 & \codepy ($110$M) & $18.04$ & $38.13$ & $63.37$ \\
 & \codepyxl & $20.50$ & $43.40$ & $56.06$ \\
 & \codepy-\cert & \bm{$31.47$}\color{red}{$\blacktriangle13.43$} & \bm{$46.42$}\color{red}{$\blacktriangle8.29$} & \bm{$66.41$}\color{red}{$\blacktriangle3.04$} \\ 
 & \ - \codepy-\cert-N & $24.91$ & $42.88$ & $54.02$ \\ 
 & \ - \codepy-\cert-NC & $19.88$ & $41.64$ & $55.82$ \\ 
 & \ - \codepy-\certg & $16.55$ & $44.07$ & $56.80$ \\ 
 \hline
\end{tabular}%
}
\caption{The pass$@k$ ($\%$) results on \peval and \neval. The absolute improvements of \cert over the base model are highlighted in \textcolor{red}{red}.
Also, we report the performance of different sketching operations (\cert-N and \cert-NC) and the performance of \certg trained for general code generation. 
}
\label{tab:table3}
\end{table}

\subsection{Closer Analysis}
We conduct some closer analyses to provide more insights.

\paragraph{Different Types of Sketching.}
As mentioned in Section~\ref{cert}, we propose three types of sketching operations. By default, \cert only anonymizes user-defined constants. The other two types include \cert-N, which anonymizes only user-defined names, and \cert-NC, which anonymizes both user-defined constants and names. As shown in Table~\ref{tab:table3}, \cert with default setting achieves the best performance. This observation may be related to the inherent characteristics of \pandas and \numpy. They are commonly used in data statistics and analysis, often involving manipulation of the data constants. Thus, it is necessary to anonymize user-defined constants. Anonymizing both user-defined constants and names would probably make the sketches too abstract.

\begin{figure}[t]
    \small
    \centering
    \includegraphics[width=0.80\linewidth]{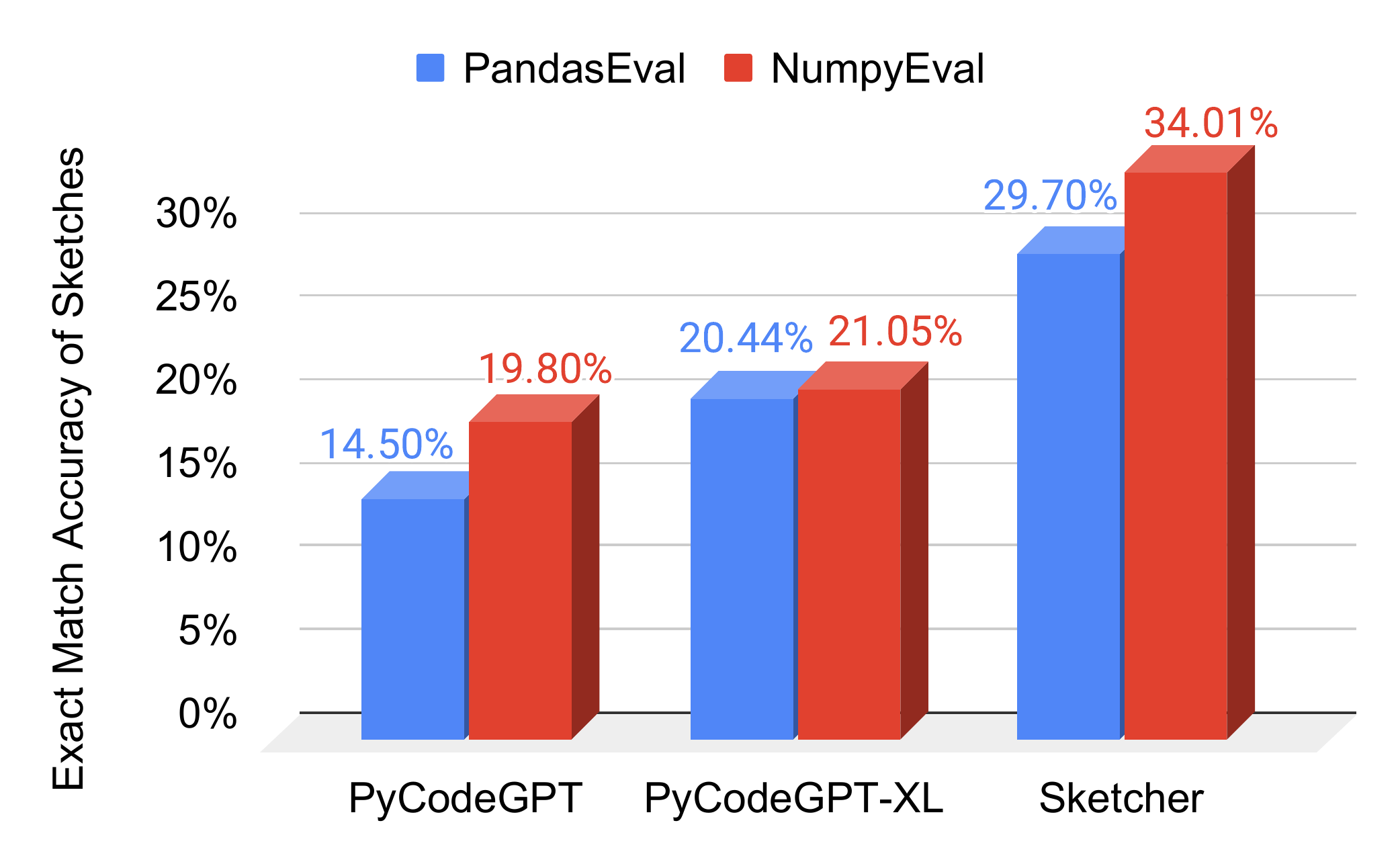}
    \caption{The exact match accuracy of sketches. The sketcher refers to the one in \codepy-\cert.
    }
    \label{fig:figure4}
\end{figure}

\paragraph{Quality of Generated Sketches.}
Intuitively, it is easier to generate a sketch than a complete code. Thus, we would like to evaluate the quality of sketches generated by the sketcher of \cert. We use exact match accuracy as the metric and include \codepy and \codepyxl for comparison. For \codepy and \codepyxl, we anonymize the user-defined constants in the predicted code to obtain the sketch. As shown in Figure~\ref{fig:figure4}, our sketcher surpasses baselines by $15.20\%$ and $14.21\%$ on \peval and \neval, respectively. It indicates that the sketcher can generate high-quality sketches, and such sketches further benefit the generator. Additionally, the generator does not necessarily require an exactly correct sketch, as the sketch is just a prompt (A case will be discussed in Section~\ref{casestudy}). 

\begin{table}[t]
\centering
\resizebox{62mm}{!}{
\begin{tabular}{l|ccc}
\hline
\textbf{Model} & \multicolumn{1}{l}{\textbf{pass$@1$}} & \multicolumn{1}{l}{\textbf{pass$@10$}} & \multicolumn{1}{l}{\textbf{pass$@100$}} \\ \hline
\codepy              & \bm{$8.33$}                        & $13.36$                         & $19.13$                                   \\
\codepy-\certg                     & $8.25$                                 & \bm{$14.12$}                                  & \bm{$20.41$}                          \\ \hline
\end{tabular}%
}
\caption{The pass$@k$ ($\%$) results of \codepy and \codepy-\certg on HumanEval.}
\label{tab:table5}
\end{table}

\begin{table}[t]
\centering
\resizebox{83mm}{!}{
\begin{tabular}{c|ll|ccc}
\hline
\textbf{Benchmark} & \textbf{Model} & \textbf{Size} & \textbf{pass$@1$} & \textbf{pass$@10$} & \textbf{pass$@100$} \\ \hline
\multirow{4}{*}{\begin{tabular}[c]{@{}c@{}}Pandas\\ Eval\end{tabular}} & \codepy-\cert & $110$M & \bm{$28.42$} & \bm{$48.04$} & $60.96$ \\
 & \codegen-\cert & $350$M & $26.40$ & $46.49$ & $58.16$ \\
 & GPT-3 & $175$B & $12.97$ & $20.54$ & $25.43$ \\
 & Codex & $12$B & $18.88$ & $43.05$ & $\mathbf{64.37}$ \\ 
 \hline
\multirow{4}{*}{\begin{tabular}[c]{@{}c@{}}Numpy\\ Eval\end{tabular}} & \codepy-\cert & $110$M & $31.47$ & $46.42$ & $66.41$ \\
 & \codegen-\cert & $350$M & $32.00$ & $49.45$ & $67.82$ \\
 & GPT-3 & $175$B & $16.25$ & $22.15$ & $27.38$ \\
 & Codex & $12$B & $\mathbf{34.42}$ & $\mathbf{55.75}$ & $\mathbf{71.74}$ \\ 
 \hline
\end{tabular}
}
\caption{GPT-3 and Codex on \peval and \neval.}
\label{tab:table7}
\end{table}

\begin{figure*}[t]
    \small
    \centering
    \includegraphics[width=0.90\linewidth]{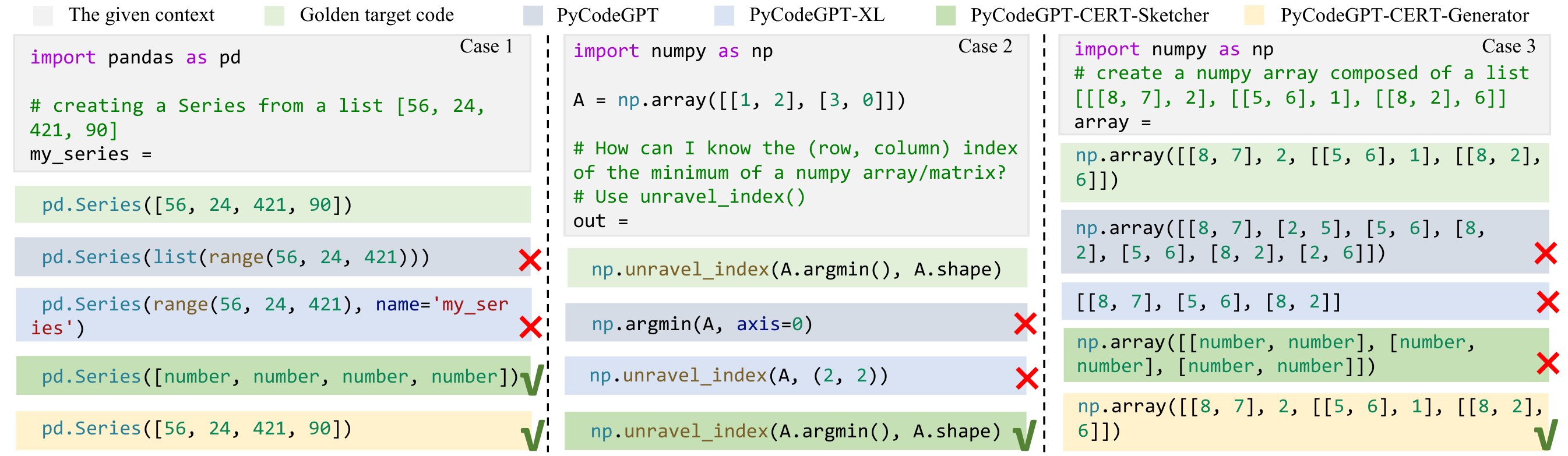}
    \caption{Three library-oriented code generation cases.}
    \label{fig:figure6}
\end{figure*}

\paragraph{\cert for General Code Generation.}
Technically speaking, \cert can also be used for general code generation tasks, not just the library-oriented ones. Concretely, following the procedure in Figure~\ref{fig:figure3}, we can continually pre-train \codepy using the whole $13.0$M python corpus instead of the extracted library-oriented files, and obtain the model we called \certg. We evaluate \certg for general code generation on HumanEval compared to the base model \codepy. As shown in Table~\ref{tab:table5}, they have similar pass$@k$ results. This observation verifies our assumption that library-oriented code snippets are more likely to share similar sketches, so it is beneficial to use sketches as prompts in this situation. But in the general case, it is not useful. 
Meanwhile the results of \certg on \peval and \neval are in Table~\ref{tab:table3}. \certg is inferior to \cert, suggesting that extracting library-oriented files is essential for \cert to learn the knowledge of library-oriented sketches.

\paragraph{Evaluation of GPT-3 and Codex.}
We evaluate GPT-3 and Codex to see how these extremely large models perform on \peval and \neval. 
As shown in Table~\ref{tab:table7}, \cert is competitive with only $110$M parameters. Such observation proves \cert's powerful code generation capability in library-oriented programming problems. 

\subsection{Case Study}\label{casestudy}
For a more comprehensive comparison, we show three cases in Figure~\ref{fig:figure6}. We show in turn the context, the golden target code, the predicted code of \codepy and \codepyxl, the sketch generated by \codepy-\cert and the predicted code of \codepy-\cert. Case $1$ is from \peval, both \codepy-\cert's sketcher and generator reach the correct results, while the baselines do not. It reveals that sketcher and generator can work well together. Case $2$ is from \neval, the sketcher predicts the correct sketch, which has no anonymous symbols, then this sketch is the final predicted code. It indicates that the sketcher has the ability to predict code without user-defined constants. At last, in Case $3$, the sketcher makes a wrong prediction {\small\texttt{pd.Series([[number*$2$]*$3$])}}, while the correct sketch is {\small\texttt{pd.Series([[number*$2$], number, [[number*$2$], number]*$2$])}}. But \codepy-\cert's generator rectifies it and finally generates the correct code. Since the sketch acts only as a prompt, it is not necessarily to be perfectly correct, which endows the generator with solid robustness.

\section{Related Work}
The most related work is the line of large pre-trained models for code. As for the encoder-style pre-trained models, they cannot be employed directly to generate code, such as CuBERT~\cite{kanade2020learning}, CodeBERT~\cite{feng2020codebert}, and GraphCodeBERT~\cite{guo2020graphcodebert}. As for the decoder-style or encoder-decoder-style ones, they are trained on large unlabelled code corpora and can work directly on code generation task, such as CodeT5~\cite{wang2021codet5}, CodeGPT~\cite{lu2021codexglue}, PLBART~\cite{ahmad-etal-2021-unified}, PolyCoder~\cite{xu2022PolyCoder}, \codegen~\cite{nijkamp2022conversational}, AlphaCode~\cite{li2022competition}, and Codex~\cite{chen2021evaluating}.
All of them focus on generating standalone code, while we investigate library-oriented code generation. Also, similar to our idea, there are several works leveraging code sketches, for example, Coarse-to-Fine~\cite{dong2018coarse}, \textsc{Bayou}~\cite{murali2017neural}, \textsc{SketchAdapt}~\cite{nye2019learning}, and \textsc{PlotCoder}~\cite{chen2021plotcoder}. 
However, they require labelled text-code paired data for fine-tuning, while our models continually pre-train on unlabelled code corpora.
For code generation benchmarks, there are few works, including APPS~\cite{hendrycks2021measuring}, HumanEval~\cite{chen2021evaluating}, and PlotCoder's dataset~\cite{chen2021plotcoder}. The former two ones focus on evaluating the capability of generating standalone code, and the last one is primarily devoted to generating plotting APIs and visualization code. \peval and \neval are dedicated to evaluating the performance of library-oriented code generation.

\section{Conclusion}
In this paper, we propose a novel approach \cert for library-oriented code generation. It leverages the code sketches and consists of a sketcher and a generator. The sketcher and generator are continually pre-trained upon a base model using unlabelled code corpora. Also, we carefully craft two benchmarks to evaluate library-oriented code generation, namely \peval and \neval. Experimental results and thorough analysis show the effectiveness of \cert. In future work, we are interested in code generation for private libraries with fewer data.

\section*{Acknowledgements}
We thank all reviewers and our colleagues, Qian Liu, Yanlin Wang and Shi Han, for their constructive comments.

\bibliographystyle{named}
\bibliography{ijcai22}

\appendix

\section{\codepy: A Democratizing Code Generation Model in Python}
Large pre-trained language models, e.g., Codex~\cite{chen2021evaluating} and AlphaCode~\cite{li2022competition}, have recently achieved surprisingly promising results on modeling source code in several programming languages. 
However, most of the state-of-the-art models are not publicly available, hindering the progress of related research topics and applications. 
To this end, we propose a publicly available code pre-trained model for Python, named \codepy, to reproduce Codex with medium size. 

\subsection{Data Construction}
It is well-known that large language models require extremely large amounts of data to exhibit its performance well.
In this section, we present the process of data collection and the data pre-processing strategies to ensure data quality.
\paragraph{Data Collection}
We first crawl $7.6$M repository pages hosted on GitHub. Then we consider the language distribution tags in each page to filter the repositories without Python files. As a result, we obtain $1.2$M Python-related repository URLs. With the filtered repository URLs, we download all the contents of each repository from GitHub. Following Codex, we remove files over $1$MB, as experienced developers usually avoid creating large source code files to maintain good readability. As a result, we get $60.6$M raw Python files under $1$MB, with a total size of $330$GB. Among these files, we further filter out duplicated files, which has been recognized as an important step by CodeParrot.
Finally, the number of unique files is reduced to $13.0$M, with a total size of $96$GB.

\paragraph{Data Pre-processing}
The data pre-processing strategies are summarized in three aspects.
\begin{itemize}
    \item According to the strategies applied to Codex and CodeParrot, we consider each source code as text and focus on \textit{line length limit} and \textit{alphanumeric rate}\footnote{It makes our model only target the English version.}. In detail, we filter the files which do not meet the four conditions: \textit{lines of code} $\ge 5$, \textit{average line length} $\le 100$, \textit{maximum line length} $\le 1000$, and \textit{alphanumeric rate} $\ge 0.98$. Note that, the fourth condition is applied after removing \textit{comments with alphanumeric rate} $< 0.5$, which is one of the following strategies.
    
    \item We also remove the automatically generated or meaningless files, for example, files with the name \textit{\_\_init\_\_.py}, \textit{setup.py} or \textit{\_pb2.py}, because these files can mislead the model during training. In addition, we remove some useless contexts from the Python files. Specifically, we remove the contexts of \textit{license description} and \textit{comments with alphanumeric rate $<$ $0.5$}, where \textit{license description} appears as a comment in the head of the code.
    \item 
    To ensure the training quality, we design two methods to perform Python syntax checking. The first method is to use Python's built-in module \verb|ast| to check the correctness of the syntax. This strategy filter out non-Python files largely, even if the file name ends with \textit{.py}. The second method applies pattern matching to leave files containing more than two typical Python keywords (e.g., \verb|def|, \verb|if|, \verb|return|, and \verb|for|).
\end{itemize}

\begin{table}[t]
	\centering
	\resizebox{70mm}{!}{
	\begin{tabular}{lc}
		\toprule
		\textbf{Hyper-parameter}                & \textbf{Value} \\ \midrule
		Learning Rate                           & $5 \times 10^{-4}$ \\ 
		Optimizer                               & ${\rm AdamW}$  \\
		Adam $\beta$                            & $0.9$, $0.95$ \\
		Adam $\epsilon$                         & $10^{-8}$ \\
		Weight Decay                            & $0.1$ \\ 
		Warmup Steps                            & $100K$ \\ 
		Learning Rate Decay                     & ${\rm Cosine}$ \\
		Batch Size (tokens)                     & $480K$ \\
		Training Steps                          & $200K~{\rm steps}, 100B~{\rm tokens}$ \\
		Context Window                          & $1024$ \\ 
		\bottomrule
	\end{tabular}
	}
	\caption{Hyper-parameters in pre-training.}
	\label{Table:TrainingHyperParameters}
\end{table}

\subsection{Model Training}
We use GPT-Neo~\cite{gpt-neo} as our base model, which is comparable to GPT-3~\cite{brown2020language} and has already been pre-trained on the Pile dataset~\cite{gao2020pile}. 
In this section, we present the details of model training, including tokenization, resampling, and hyper-parameters.

\paragraph{Tokenization}
The tokenizer of GPT-Neo models is based on GPT-2~\cite{radford2019language} tokenizer, which applies the byte-level version of Byte Pair Encoding (BPE)~\cite{sennrich2016neural}.
The tokenizer is trained on the Pile dataset with a vocabulary of $50$K. 
Since the distribution of source code words differ from that of natural text, the original GPT-Neo tokenizer is not very effective for encoding Python source code. 
Thus, we follow CodeParrot to train a new byte-level BPE tokenizer from scratch on our collected code data. 
Finally, we set the vocabulary size to $32$K, which allows us to encode code using approximately 40\% fewer tokens. 

\paragraph{Resampling}
Since different files have different importance for training, 
we design a data resampling strategy to make high-quality files appear more often, while keeping a balance to make each file appear at least once throughout the training process. 
In detail, we evaluate the quality of a Python file in two aspects: repository star count and unit test function rate. The repository star count is determined by GitHub users and is the main factor in measuring repository quality. We do not use the star count directly for filtering data because most files have very few stars.
The unit test function rate is the number of unit test functions divided by the number of functions. We introduce it to reduce the weight of files used for testing, because test files often contain a lot of user-defined numeric constants or string constants.

\paragraph{Hyper-parameters}
\codepy shares the same configurations with original GPT-Neo 125M model except for the vocabulary size. It is trained on $16$ V$100$ ($32$GB) GPUs for about $2$ days. Table~\ref{Table:TrainingHyperParameters} lists the hyper-parameters.

\begin{table}[t]
	\centering
	\resizebox{85mm}{!}{
	\begin{tabular}{crrrr}
	    \toprule
		\multirow{2}{*}{\textbf{Model}} & \multirow{2}{*}{\textbf{Params}} & \multicolumn{3}{c}{\textbf{pass$@k$}}\\
		\cmidrule(lr){3-5}
		& & \textit{k=$1$} & \textit{k=$10$} & \textit{k=$100$} \\ 
		\midrule
		\multicolumn{5}{c}{\textit{Parameter Scale:  ${\sim}100$M}} \\
		{TabNine}  & -- & $2.58$\%    & $4.35$\%    & $7.59$\% \\
		GPT-Neo   & $125$M & $0.75$\%    & $1.88$\%    & $2.97$\% \\
		CodeParrot   & $110$M & $3.80$\%     & $6.57$\%    
		& $12.78$\% \\
		PolyCoder & $160$M & $2.13$\%    & $3.35$\%    & $4.88$\% \\ 
        Codex    & $85$M & $8.22$\%    & $12.81$\%   & $\mathbf{22.40}$\% \\ 
		AlphaCode & $89$M & $4.3$\%     & $12.2$\%    & $20.0$\% \\
	    \codepy (Ours) & $110$M & {$\mathbf{8.33}$\%} & {$\mathbf{13.36}$\%}   & $19.13$\% \\
		\midrule
		\multicolumn{5}{c}{\textit{Parameter Scale:  ${\sim}500$M}} \\
		PolyCoder  & $400$M & $2.96$\%    & $5.29$\%    & $11.59$\% \\ 
		Codex      & $300$M & $13.17$\%   & $20.37$\%   & $36.27$\% \\
		Codex       & $679$M & $\mathbf{16.22}$\%   & $\mathbf{25.7}$\%    & $\mathbf{40.95}$\% \\
		AlphaCode   & $302$M & $11.6$\%    & $18.8$\%    & $31.8$\% \\
		AlphaCode & $685$M & $14.2$\%    & $24.4$\%    & $38.8$\% \\
		\textsc{CodeGen-Mono} & $350$M & $12.76$\%    & $23.11$\%    & $35.19$\% \\
        \midrule
        \multicolumn{5}{c}{\textit{Parameter Scale:  ${\sim}1$B}} \\
		GPT-Neo      & $1.3$B & $4.97$\%    & $7.47$\%    & $16.30$\% \\
		CodeParrot  & $1.5$B & $3.58$\%    & $8.03$\%    & $14.96$\% \\ 
        AlphaCode  & $1.1$B & $\mathbf{17.1}$\%    & $\mathbf{28.2}$\%    & $\mathbf{45.3}$\% \\ 
		\midrule
		\multicolumn{5}{c}{\textit{Parameter Scale:  ${\sim}2$B}} \\
		GPT-Neo        & $2.7$B & $6.41$\%    & $11.27$\%   & $21.37$\% \\
		PolyCoder       & $2.7$B & $5.59$\%    & $9.84$\%    & $17.68$\% \\ 
        Codex       & $2.5$B & $21.36$\%   & $35.42$\%   & $\mathbf{59.50}$\% \\
	    \textsc{CodeGen-Mono} & $2.7$B & $\mathbf{23.70}$\%    & $\mathbf{36.64}$\%    & $57.01$\% \\
        \midrule
        \multicolumn{5}{c}{\textit{Parameter Scale:  ${\sim}6$B}} \\
        GPT-J  & $6$B & $11.62$\%   & $15.74$\%   & $27.74$\% \\ 
        \textsc{CodeGen-Mono} & $6.1$B & $\mathbf{26.13}$\%    & $\mathbf{42.29}$\%    & $\mathbf{65.82}$\% \\
        \midrule
        \multicolumn{5}{c}{\textit{Parameter Scale:  \textgreater$10$B}} \\
        Codex        & $12$B & $28.81$\%   & $46.81$\%   & $72.31$\% \\ 
        \textsc{CodeGen-Mono} & $16.1$B & $\mathbf{29.28}$\%    & $\mathbf{49.86}$\%    & $\mathbf{75.00}$\% \\
		\bottomrule
	\end{tabular}
	}
	\caption{Experimental results of models under different parameter scales on the HumanEval Dataset. For AlphaCode, we report the pre-trained decoder-only results from their paper. For TabNine, we do know the parameter scale, so we put it to the first scale group.}
	\label{Table:HumanEval}
\end{table}

\subsection{Experiments}
We conduct experiments on HumanEval and CodeXGLUE to show the effectiveness of \codepy.
\paragraph{Results on HumanEval}
HumanEval~\cite{chen2021evaluating} contains $164$ hand-written code generation problems. As mentioned in Section~\ref{experiments}, the evaluation metric is pass$@k$. We use the same parameters as those used by Codex, except for the stop sequences. The stop sequences include \verb|\nclass|, \verb|\ndef|, \verb|\n#|, \verb|\n@|, \verb|\nprint|, and \verb|\nif|. We generate $200$ programs and apply nucleus sampling using $p = 0.95$. As in previous work, we try various temperatures (from $0.1$ to $1.0$ with the interval of $0.1$) and report the best result for each $k$. 

Table~\ref{Table:HumanEval} shows the experimental results.
Even though the original GPT-Neo models ($125$M, $1.3$B and $2.7$B) and GPT-J $6$B~\cite{gpt-j} are trained on the dataset containing GitHub files, they do not achieve satisfactory performance on HumanEval compared to Codex. For the open-source CodeParrot models ($110$M and $1.5$B), as we mentioned before, they outperform the corresponding ones of GPT-Neo but still not enough to compare with Codex. AlphaCode~\cite{li2022competition} also evaluates their decoder-only model on HumanEval, and the performance is slightly worse than Codex. PolyCoder~\cite{xu2022PolyCoder} is pre-trained on several programming languages, and it shows even worse performance than CodeParrot. However, our pre-trained $110$M model can obtain comparable performance to Codex $85$M and outperform CodeParrot $110$M by $4.53$\% on pass$@1$, $6.79$\% on pass$@10$ and $6.35$\% on pass$@100$. During our paper review, \codegen~\cite{nijkamp2022conversational} also provided code generation models of various sizes from $350$M to $16.1$B and obtained good performance. We omit CodeT5, CodeGPT, and CodeClippy, since they all get $0\%$ on pass$@1$, pass$@10$, and pass$@100$.
\begin{table}[t]
	\centering
	\resizebox{87mm}{!}{
	\begin{tabular}{ccccc}
	    \toprule
		\multirow{2}{*}{\textbf{Model}} & \multirow{2}{*}{\textbf{Params}} & \textbf{Token Level} & \multicolumn{2}{c}{\textbf{Line Level}} \\
		\cmidrule(lr){3-3}
        \cmidrule(lr){4-5}
		 &  & Accuracy & EM & ES \\
		\midrule
		LSTM                   & -- & $58.00$\%    & $17.93$\%    & $50.05$\% \\
		Transformer            & -- & $73.26$\%    & $36.65$\%    & $67.51$\% \\
		GPT-2                  & $117$M & $74.22$\%    & $38.55$\%    & $68.94$\% \\
		CodeGPT                & $124$M & $74.93$\%    & $39.11$\%    & $69.69$\% \\
		CodeGPT-Adapted        & $124$M & $75.11$\%    & $39.65$\%    & $69.84$\% \\
		CodeParrot        & $110$M & $77.22$\%    & $42.10$\%    & $71.07$\% \\
		\midrule
		\codepy (Ours)         & $110$M & ${\mathbf{80.24}}$\%    & {$\mathbf{44.77}$}\%   & {$\mathbf{73.09}$\%} \\
		\bottomrule
	\end{tabular}
	}
	\caption{Experimental results on CodeXGLUE Code Completion task dataset PY150. Results of LSTM, Transformer, GPT-2, CodeGPT and CodeGPT-Adapted are from \protect\citet{lu2021codexglue}. Exact match and edit similarity are abbreviated as EM and ES.}
	\label{Evaluation:CodeXGLUEDataset}
\end{table}
\paragraph{Results on CodeXGLUE}
CodeXGLUE~\cite{lu2021codexglue} is a benchmark for programming language understanding and generation tasks. In total, CodeXGLUE provides datasets for 14 tasks, in which the dataset PY150 is for code completion task. Table~\ref{Evaluation:CodeXGLUEDataset} shows the experimental results. Note that we fine-tuned \codepy and CodeParrot on PY150 training dataset. The results show that CodeGPT is worse than the fine-tuned CodeParrot. However \codepy wins the CodeParrot with $3.02$\% higher score on token level completion accuracy. On line level completion, \codepy also achieves $2.67$\% and $2.02$\% higher accuracy on exact match and edit similarity, respectively.

\end{document}